\documentclass[final,5p,times,twocolumn]{elsarticle}   
\usepackage{graphicx}
\graphicspath{{./}{../results/}}   
\usepackage{amsmath}
\usepackage{booktabs}
\usepackage{xcolor}
\usepackage[colorlinks=true,allcolors=blue]{hyperref}

\newcommand{\kzs}{{\mathrm{K}^0_\mathrm{S}}}
\newcommand{\lam}{\Lambda}
\newcommand{\dndeta}{\mathrm{d}N_\mathrm{ch}/\mathrm{d}\eta}
\newcommand{\ntrk}{N_\mathrm{trk}^\mathrm{offline}}
\newcommand{\sperp}{S_\perp}

\journal{Physics Letters B}

\begin{document}

\begin{frontmatter}

\title{Multiplicity and transverse density as ordering variables of the
$\lam/\kzs$ ratio in oxygen--oxygen collisions at the LHC}

\author[iitm]{Neeraj\corref{cor1}}
\ead{neeraj.neeraj@cern.ch}
\author[iitm]{Amal Sarkar}
\ead{amal.sarkar@cern.ch}
\cortext[cor1]{Corresponding author}
\affiliation[iitm]{organization={Indian Institute of Technology Mandi},
            city={Mandi},
            postcode={175075},
            state={Himachal Pradesh},
            country={India}}

\begin{abstract}
Baryon-to-meson enhancements are commonly compared across collision systems at equal multiplicity. Using published data only, we test whether the $p_\mathrm{T}$ dependence of the CMS $\lam/\kzs$ ratio in minimum-bias O--O collisions at $\sqrt{s_\mathrm{NN}}=5.36$~TeV can be obtained from CMS multiplicity-differential p--Pb and Pb--Pb measurements weighted with the O--O multiplicity distribution. The prediction from p--Pb collisions of equal multiplicity is harder than the data, with a tilt of $+0.35\pm0.04$ in $\ln(\mathrm{prediction}/\mathrm{data})$ versus $\ln p_\mathrm{T}$, whereas that from peripheral Pb--Pb collisions is consistent with the data. In a Glauber model calibrated on O--O data, O--O and peripheral Pb--Pb collisions of equal multiplicity have similar transverse densities $\rho=\langle(\dndeta)/\sperp\rangle$, about half of that in p--Pb. Matching the p--Pb reference in $\rho$ instead of multiplicity reduces the tilt by about 60\%, to $+0.14\pm0.04$, with nucleons as sources; the tension is fully removed ($-0.02\pm0.04$) only when sub-nucleonic structure is included (three constituents per nucleon), so the conclusion depends on the modeled size of high-multiplicity p--Pb collisions. The published blast-wave velocities of p--Pb and Pb--Pb collisions, which differ at equal multiplicity, are close at equal $\rho$. Multiplicity alone thus does not determine the $\lam/\kzs$ shape across systems; the data are consistent with a radial boost governed by the transverse density, as expected from hydrodynamics. The same pattern is found in EPOS4, which has a hydrodynamic core, but not in PYTHIA Angantyr. For Ne--Ne collisions, which have the transverse density of O--O at equal multiplicity, the minimum-bias ratio is predicted to agree with O--O within 3--5\%, as EPOS4 also predicts.
\end{abstract}


\end{frontmatter}

\section{Introduction}
\label{sec:intro}

In collisions of heavy nuclei the baryon-to-meson ratios, such as $\mathrm{p}/\pi$ and $\lam/\kzs$, are enhanced at intermediate transverse momentum, $2\lesssim p_\mathrm{T}\lesssim 5$~GeV, relative to proton--proton collisions~\cite{ALICE:2013cdo}. The enhancement is commonly attributed to collective radial flow, which pushes heavier hadrons to higher $p_\mathrm{T}$, and to hadronization by quark recombination~\cite{Fries:2003vb,Greco:2003xt,Hwa:2002tu}. Similar features, together with an enhancement of strange-hadron yields, are observed in high-multiplicity pp and p--Pb collisions~\cite{ALICE:2013wgn,CMS:2016zzh,ALICE:2018pal,ALICE:2016fzo,ALICE:2019avo}, where they evolve smoothly with the charged-particle multiplicity~\cite{Nagle:2018nvi,Schenke:2021mxx}. Multiplicity has thus become the most widely used variable for comparing small and large systems.

For observables driven by collective expansion, however, the system's transverse size is expected to matter as well. Elliptic flow has long been discussed in terms of the multiplicity per unit transverse area, $(1/S)\,\mathrm{d}N/\mathrm{d}y$~\cite{Voloshin:1999gs,Bhalerao:2005mm,Drescher:2007cd}, and in hydrodynamics the mean transverse momentum is governed by the entropy density rather than the total entropy~\cite{Gardim:2019xjs}. High-multiplicity p--Pb collisions, being more compact than Pb--Pb collisions of the same multiplicity, were predicted to develop a stronger radial flow~\cite{Bozek:2013df,Bozek:2013ska}; the mean transverse momentum~\cite{ALICE:2013rdo} and the blast-wave velocities~\cite{CMS:2016zzh} measured in p--Pb collisions are indeed larger than in Pb--Pb collisions of equal multiplicity. How compact p--Pb collisions are depends on the initial-state geometry, in particular on sub-nucleonic fluctuations~\cite{Bzdak:2013zma,Welsh:2016siu,Mantysaari:2016ykx,Moreland:2018gsh}. In core--corona approaches, the fraction of hadrons produced in a dense core likewise depends on the local density rather than on the total multiplicity~\cite{Werner:2007bf,Becattini:2008ya,Aichelin:2008mi,Kanakubo:2019ogh}.

Collisions of light nuclei at the LHC~\cite{Brewer:2021kiv,CMS:2026sai,CMS:2026aie} make it possible to separate the two variables with data. Minimum-bias O--O collisions reach the multiplicities of high-multiplicity p--Pb collisions, but with a transverse overlap region similar to that of peripheral Pb--Pb collisions. CMS has measured a pronounced $\lam/\kzs$ enhancement in minimum-bias O--O collisions at $\sqrt{s_\mathrm{NN}}=5.36$~TeV~\cite{CMS:2026aie}, and preliminary ALICE results indicate that the mean transverse momenta of strange hadrons in O--O collisions follow the Pb--Pb rather than the p--Pb trend~\cite{Pucillo:2026wdz}. The Ne--Ne collisions recorded in the same run, with the strongly deformed $^{20}$Ne nucleus, add a second light-ion system~\cite{Giacalone:2024luz}.

In this Letter we ask quantitatively, using only published data, whether the $p_\mathrm{T}$ shape of the minimum-bias O--O $\lam/\kzs$ ratio can be obtained from p--Pb or Pb--Pb collisions of equal multiplicity (Sec.~\ref{sec:mixture}) or of equal transverse density computed with a Glauber model (Sec.~\ref{sec:density}). The same procedure is applied to EPOS4 and PYTHIA Angantyr events (Sec.~\ref{sec:models}), and the Ne--Ne ratio is predicted (Sec.~\ref{sec:nene}). Further details are given in the Supplementary Material, cited below as Sec.~S1--S6.

\section{Data and method}
\label{sec:method}

All inputs are public HEPData records (list in Sec.~S1): the CMS minimum-bias O--O $\kzs$ and $\lam$ spectra at $|y|<1$ and the pp reference at the same energy~\cite{CMS:2026aie}, the O--O $\dndeta$ in 5\% centrality classes~\cite{CMS:2026sai}, and the CMS $\kzs$ and $\lam+\bar\lam$ spectra and blast-wave parameters in classes of the number of reconstructed tracks $\ntrk$ ($|\eta|<2.4$, $p_\mathrm{T}>0.4$~GeV) for p--Pb at 5.02~TeV, pp at 7~TeV and Pb--Pb at 2.76~TeV~\cite{CMS:2016zzh}. All $\lam$ yields are corrected for feed-down from $\Xi$ and $\Omega$ decays. The central analysis uses only CMS data, since comparisons across experiments are sensitive to conventions (Sec.~S1).

Minimum-bias O--O events are a mixture of multiplicities. If the spectra depended on multiplicity only, the minimum-bias ratio would be
\begin{equation}
R_\mathrm{MB}(p_\mathrm{T}) =
\frac{\sum_c \lam_\mathrm{ref}(p_\mathrm{T}; N_c)}
     {\sum_c \kzs\!\,_\mathrm{ref}(p_\mathrm{T}; N_c)},
\label{eq:mixture}
\end{equation}
where $c$ runs over twenty 5\% O--O centrality classes, $N_c=\kappa\,(\dndeta)_c$ is the corresponding $\ntrk$ with $\kappa=2.75\pm0.35$, and the reference per-event spectra are interpolated between the measured classes linearly in $(\log N,\log \mathrm{yield})$. The 70--100\% classes are extrapolated exponentially; beyond the highest reference class, either a log-linear extrapolation or a saturated shape is used.

Since the reference spectra come from an earlier CMS analysis, each prediction is calibrated with the double ratio
\begin{equation}
s_\mathrm{cal}(p_\mathrm{T}) =
\frac{[\lam/\kzs]_\mathrm{pp,MB}^\mathrm{2026,\,5.36\,TeV}}
     {\sum_c f_c\,\lam_c(p_\mathrm{T};rN_c) \big/ \sum_c f_c\,\kzs\!\,_c(p_\mathrm{T};rN_c)},
\label{eq:cal}
\end{equation}
whose denominator is the minimum-bias pp ratio of the earlier analysis, built like Eq.~(\ref{eq:mixture}) from its six 7~TeV pp classes with their fractions $f_c$ of the minimum-bias sample~\cite{CMS:2016zzh} and multiplicities scaled by $r=0.93$, the measured energy dependence of the charged-particle density~\cite{ALICE:2022kol}. The factor $s_\mathrm{cal}$ is 0.94 at 1~GeV and 0.81--0.87 above 1.3~GeV; it changes the tilt defined below by $-0.05$. The uncalibrated predictions and the calibration with $r=1$ are included in the variations.

Data $D$ and prediction $P$ are compared with $\chi^2(a)=(D-aP)^\mathrm{T}(C_D+a^2C_P)^{-1}(D-aP)$, profiled over the free scale $a$, so that only the $p_\mathrm{T}$ shape is tested. Since HEPData provides only total systematic uncertainties, the covariances are built by decomposing each systematic uncertainty into normalization-like, linear-in-$\log p_\mathrm{T}$, low-$p_\mathrm{T}$, and uncorrelated components, from fully uncorrelated to almost fully correlated; a ``balanced'' decomposition is the nominal one (Sec.~S3). As a single measure of a shape difference, we fit
\begin{equation}
\ln\frac{P(p_\mathrm{T})}{D(p_\mathrm{T})}=\ln a+s\,\ln\frac{p_\mathrm{T}}{2~\mathrm{GeV}}
\label{eq:tilt}
\end{equation}
by generalized least squares with the same covariances. The tilt $s$ is insensitive to normalization; $s>0$ means the prediction is harder than the data, as expected for a stronger radial boost.

\section{Is multiplicity sufficient?}
\label{sec:mixture}
\begin{figure}[t]
\includegraphics[width=\columnwidth]{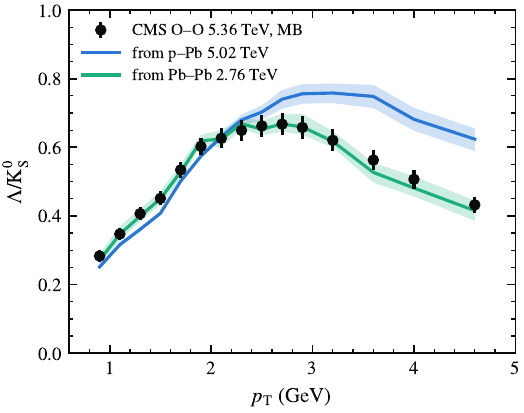}
\caption{Minimum-bias O--O $\lam/\kzs$~\cite{CMS:2026aie} compared with the multiplicity-only predictions of Eq.~(\ref{eq:mixture}) from CMS p--Pb and Pb--Pb multiplicity classes~\cite{CMS:2016zzh}. Bands: variations of $\kappa$ and of the extrapolation, each with the statistical uncertainty of the reference classes (bootstrap, 68\% interval). Each prediction is multiplied by its best-fit scale $a$ of the shape test (balanced correlation model; 1.14 for p--Pb, 0.98 for Pb--Pb).}
\label{fig:mixture}
\end{figure}

Figure~\ref{fig:mixture} compares the measured O--O ratio with the predictions. The Pb--Pb-based prediction follows the data over the full $p_\mathrm{T}$ range. The prediction based on energy-matched p--Pb data lies below the data for $p_\mathrm{T}<2$~GeV and above it for $p_\mathrm{T}>2.2$~GeV, increasingly so at high $p_\mathrm{T}$, the pattern expected from a stronger radial boost. In the shape test it is disfavored, with $\chi^2/\mathrm{ndf}=29$--$64/14$ ($p\le0.01$) for uncorrelated uncertainties and $69$--$125/14$ ($p<3\times10^{-9}$) for the balanced correlation model, whereas the Pb--Pb prediction gives $\chi^2/\mathrm{ndf}=1.0$--$19.5/14$ over all variations. In terms of the tilt, the p--Pb prediction is harder than the data by $s=+0.35\pm0.04$ ($+0.29$ to $+0.43$ over all variations, $8.7\sigma$), and the Pb--Pb prediction gives $s=+0.01\pm0.04$.

The Pb--Pb reference is corrected from 2.76 to 5.02~TeV with the change of the $\lam/\kzs$ shape at matched $\dndeta$ measured by ALICE~\cite{ALICE:2013cdo,ALICE:2025cqy}. This factor deviates from unity by at most 6.5\% (rms 3.4\%) and shifts the Pb--Pb tilt by $+0.03$. Scaled in $\ln\sqrt{s_\mathrm{NN}}$ to the remaining step from 5.02 to 5.36~TeV, which also separates the p--Pb reference from the O--O data, it changes the tilts by less than 0.005 (Sec.~S1).

Three of the twenty O--O classes (0--15\%, 42\% of the charged-particle yield) lie beyond the highest p--Pb class ($\ntrk\approx230$). Built from the seventeen interpolated classes alone, the p--Pb prediction is still harder than the data, $s=+0.24\pm0.04$, although it omits the most central classes, whose spectra are the hardest; the three extrapolated classes alone give $s=+0.53$ ($+0.47$ with the saturated shape). The extrapolation therefore increases the tension but does not create it. A prediction from the 7~TeV pp classes is not used in the tests, because most of its weight comes from classes beyond the measured pp multiplicity range (Sec.~S2).

\section{Transverse density as an alternative ordering variable}
\label{sec:density}

We use a Monte Carlo Glauber model~\cite{Miller:2007ri,Loizides:2017ack} in which each participant and each binary collision is a source of Gamma-distributed weight, with the fraction $f=0.5$ fixed by the O--O centrality dependence of $\dndeta$~\cite{CMS:2026sai}, which is reproduced to 2--3\% (rms). The transverse area $\sperp=4\pi\sqrt{\sigma_x^2\sigma_y^2-\sigma_{xy}^2}$ is computed from the source weights with a Gaussian nucleon width of 0.5~fm, and $\rho=\langle(\dndeta)/\sperp\rangle$. Variations of $f$, of the fluctuations and of the nucleon width, and a version with three constituents per nucleon~\cite{Loizides:2016djv} with Gaussian spread $\sigma_c=0.4$ or 0.55~fm, are used as systematic uncertainties (Sec.~S4). At $\ntrk\approx130$ we find $\rho=6.9$~fm$^{-2}$ in p--Pb and 3.9~fm$^{-2}$ in both peripheral Pb--Pb and O--O collisions. The absolute scale of $\rho$ depends on the source width, but the ratio between p--Pb and Pb--Pb at equal multiplicity is robust (1.8 without and 2.1 with sub-nucleonic structure). With constituents, the density of peripheral Pb--Pb classes becomes nearly independent of multiplicity, so density matching to Pb--Pb is ill-conditioned, and the p--Pb reference provides the more sensitive test.

\begin{figure*}[t]
\includegraphics[width=\textwidth]{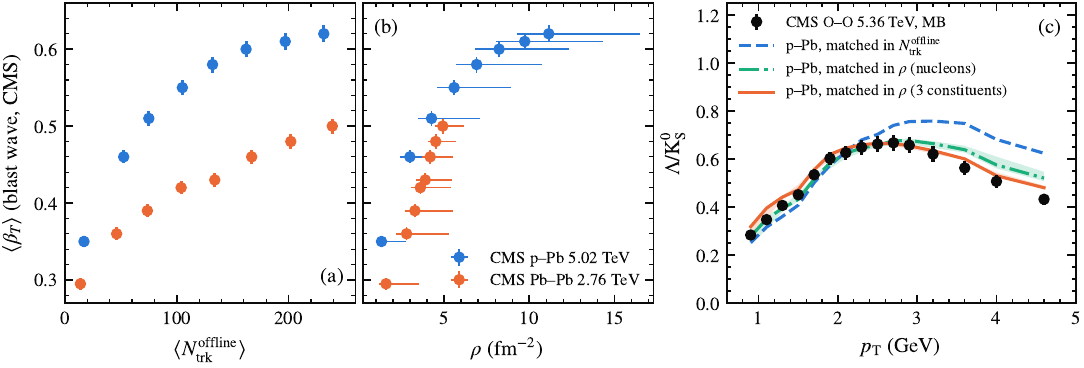}
\caption{(a), (b): blast-wave $\langle\beta_T\rangle$ published by CMS~\cite{CMS:2016zzh} for p--Pb and Pb--Pb multiplicity classes versus $\ntrk$ and versus the Glauber transverse density (horizontal bars: Glauber variations). (c): O--O minimum-bias $\lam/\kzs$ compared with the p--Pb-based prediction matched in multiplicity (dashed) and in density with nucleon sources (dash-dotted) or three constituents per nucleon (solid); bands: Glauber variations of the same source model, extrapolation and calibration. Each prediction is multiplied by its best-fit scale (balanced correlation model; 1.14, 1.11 and 1.17). With $p_\mathrm{T}$-correlated uncertainties the best fit need not pass through the points: the deviations of the constituent prediction at the lowest and highest $p_\mathrm{T}$ are covered by the correlated components ($\chi^2/\mathrm{ndf}=11.4/14$).}
\label{fig:density}
\end{figure*}

The blast-wave parameters published by CMS separate p--Pb and Pb--Pb strongly at equal multiplicity (0.58 versus 0.43 at $\ntrk\approx130$), but lie close to a common curve as a function of $\rho$ [Fig.~\ref{fig:density}(a),(b)].

Replacing $N_c$ in Eq.~(\ref{eq:mixture}) by the O--O class density $\rho_c$, taking the reference spectra per unit $\dndeta$ at the same $\rho$ and rescaling them to the O--O class yield, gives the density-matched prediction [Fig.~\ref{fig:density}(c)]. For p--Pb, the shape $\chi^2$ decreases by 25--85 units. With nucleons as sources, the prediction becomes compatible with the data for uncorrelated uncertainties ($p=0.15$--$0.99$), but a tension remains for the balanced correlation model ($p=9\times10^{-4}$--$0.08$). With sub-nucleonic structure, which makes high-multiplicity p--Pb collisions more compact, it agrees with the data: $\chi^2/\mathrm{ndf}=2.9$--$4.2/14$ for uncorrelated and $11.4$--$15.9/14$ ($p=0.3$--$0.65$) for balanced correlations. Only the almost fully correlated decomposition still disfavors it, as it also disfavors the uncalibrated Pb--Pb prediction.

\begin{table}[t]
\caption{Tilt $s$ of Eq.~(\ref{eq:tilt}) for predictions of the minimum-bias O--O $\lam/\kzs$ ratio (balanced correlation model); range: $\kappa$, extrapolation and calibration variations.}
\label{tab:tilt}
\centering
\setlength{\tabcolsep}{3.5pt}
\begin{tabular}{lcc}
\toprule
Prediction & $s$ & range \\
\midrule
p--Pb, $\ntrk$ & $+0.35\pm0.04$ & $+0.29$\,\dots\,$+0.43$\\
p--Pb, $\rho$, nucleons & $+0.14\pm0.04$ & $+0.12$\,\dots\,$+0.19$\\
p--Pb, $\rho$, constituents & $-0.02\pm0.04$ & $-0.04$\,\dots\,$+0.03$\\
Pb--Pb, $\ntrk$ (energy corr.) & $+0.01\pm0.04$ & $-0.05$\,\dots\,$+0.09$\\
\bottomrule
\end{tabular}
\end{table}

The tilt summarizes this progression (Table~\ref{tab:tilt}): density matching removes about 60\% of the tilt with nucleons as sources and all of it with sub-nucleonic structure, $s=-0.02\pm0.04$. In all variations of the procedure, the multiplicity-matched tilt stays above $+0.29$ and the density-matched tilt with constituents within $\pm0.04$ of zero; the dominant uncertainty of the density picture is the modeling of the p--Pb transverse size (Sec.~S3, Table~S1).

\section{Model comparison}
\label{sec:models}
To test whether the density ordering emerges from known dynamics, we apply the same procedure to generator events: EPOS~4.0.3~\cite{Werner:2023zvo}, with core--corona separation, 3+1D viscous hydrodynamics for the core and the UrQMD afterburner~\cite{Bass:1998ca,Bleicher:1999xi}, and PYTHIA~8.317 Angantyr~\cite{Bierlich:2018xfw,Bierlich:2022pfr}, without hydrodynamic expansion, with the default settings, with rope hadronization and string shoving~\cite{Bierlich:2014xba,Bierlich:2017vhg}, and with QCD-based color reconnection. Event classes, acceptance, and multiplicity estimators follow the data. Within each model, the minimum-bias O--O ratio is predicted from the model's own p--Pb classes, matched in $\ntrk$ or in $\rho$ with the same Glauber tables as for the data, and the tilt is computed with bootstrap uncertainties (Sec.~S5).

\begin{figure}[t]
\includegraphics[width=\columnwidth]{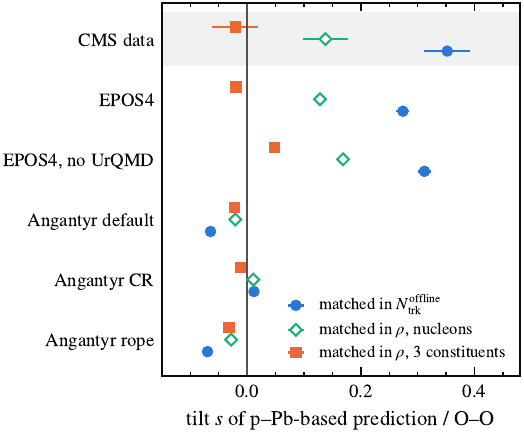}
\caption{Tilt $s$ of Eq.~(\ref{eq:tilt}) of the p--Pb-based prediction of the minimum-bias O--O $\lam/\kzs$ ratio relative to the O--O ratio of the same data set or model, for matching in multiplicity (circles) and in Glauber transverse density with nucleon (open diamonds) or constituent (squares) sources. Data: balanced correlation model; models: statistical (bootstrap) uncertainties.}
\label{fig:tilt}
\end{figure}

EPOS4 reproduces the pattern found in the data (Fig.~\ref{fig:tilt}): matched in multiplicity, the p--Pb-based prediction is too hard ($s=+0.274\pm0.011$; data $+0.35\pm0.04$); matched in density with nucleon sources the tilt is reduced to $+0.128\pm0.007$ (data $+0.14\pm0.04$), and with constituent sources it vanishes, $s=-0.020\pm0.009$ (data $-0.02\pm0.04$). In EPOS4, the effect is thus related to the geometry of the hydrodynamic core: at equal multiplicity, the compact p--Pb core expands more strongly than the O--O core. Without the hadronic cascade, the tilts are $+0.312\pm0.012$ and $+0.049\pm0.009$, so the ordering arises before the hadronic phase. Angantyr, in contrast, shows no sensitivity to the transverse geometry: the tilts are small for both matchings ($-0.064\pm0.008$ and $-0.022\pm0.004$ for the default settings; similar for the other variants), and are opposite in sign to the data for multiplicity matching. In absolute terms, neither model describes the measured ratio in detail (Sec.~S5, Fig.~S2): EPOS4 reproduces its normalization and the position of the maximum but underestimates it by up to 30\% below 1.5~GeV, whereas Angantyr underestimates the enhancement by about 40\% with the default settings.

\section{Prediction for Ne--Ne collisions}
\label{sec:nene}
\begin{figure*}[t]
\includegraphics[width=\textwidth]{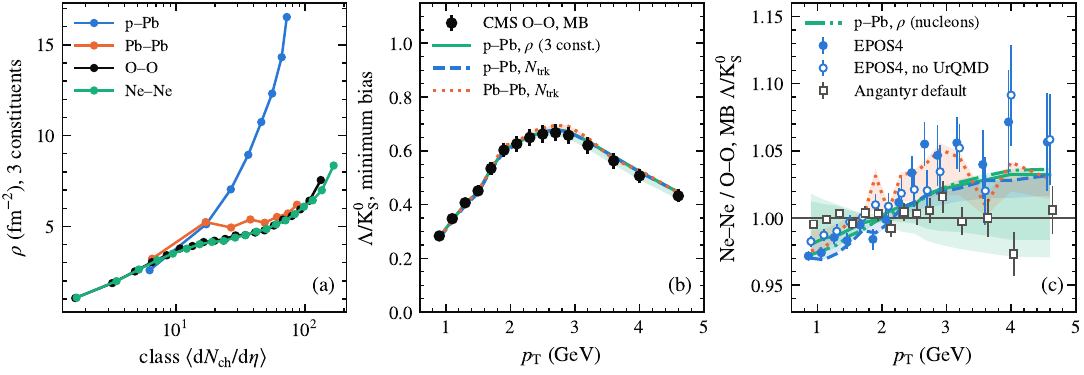}
\caption{(a) Glauber transverse density (three constituents per nucleon) of the p--Pb, Pb--Pb, O--O and Ne--Ne event classes versus their $\dndeta$. (b) Predicted minimum-bias Ne--Ne $\lam/\kzs$ at $\sqrt{s_\mathrm{NN}}=5.36$~TeV, anchored to the CMS O--O data~\cite{CMS:2026aie} (points) as in Eq.~(\ref{eq:nene}), for matching to p--Pb in $\rho$ (solid) or $\ntrk$ (dashed) and to energy-corrected Pb--Pb in $\ntrk$ (dotted). (c) The double ratio $P_\mathrm{NeNe}/P_\mathrm{OO}$ of each method (bands: Glauber, $\ntrk$-scale and extrapolation variations, including the $^{20}$Ne shape), compared with the Ne--Ne/O--O ratio of the minimum-bias $\lam/\kzs$ in EPOS4 with (full circles) and without (open circles) the hadronic cascade and in default PYTHIA Angantyr (open squares); statistical uncertainties.}
\label{fig:nene}
\end{figure*}

The 2025 LHC light-ion run also delivered Ne--Ne collisions at $\sqrt{s_\mathrm{NN}}=5.36$~TeV. The $^{20}$Ne nucleus is larger than $^{16}$O and strongly deformed, so it is natural to ask whether it probes a different transverse density at the same multiplicity. No strange-hadron measurement exists yet for Ne--Ne; we therefore predict its minimum-bias $\lam/\kzs$ with the methods tested above. $^{20}$Ne is described as a deformed Woods--Saxon distribution with the parameters of TGlauberMC~v3.3~\cite{Loizides:2025ule}, with a more strongly deformed and a spherical nucleus as variations. The per-source yield is taken from the O--O calibration, so that the Ne--Ne centrality dependence of $\dndeta$ is a prediction; it agrees within a few percent with the Ne--Ne/O--O ratio measured by ATLAS~\cite{ATLAS:2026zgq}, and the two deformed parameterizations bracket the data (Sec.~S4). At equal multiplicity, the two systems have almost the same transverse density [Fig.~\ref{fig:nene}(a)]: the Ne--Ne/O--O ratio of $\rho$ is 0.97--1.01 for $20\le\dndeta\le100$ with the deformed nucleus and 0.89--0.96 with the spherical one, because a given multiplicity is reached in more peripheral Ne--Ne than O--O collisions, with overlap regions of similar size.

Each method of Secs.~\ref{sec:mixture} and \ref{sec:density} is applied to the Ne--Ne classes and anchored to the measured O--O ratio,
\begin{equation}
\left[\lam/\kzs\right]_\mathrm{NeNe} =
\left[\lam/\kzs\right]_\mathrm{OO}^\mathrm{CMS}\times
\frac{P_\mathrm{NeNe}}{P_\mathrm{OO}},
\label{eq:nene}
\end{equation}
where $P$ is the prediction of the same method for each system, so that method-specific calibrations cancel. All methods give double ratios between 0.97 and 1.04 over $0.8<p_\mathrm{T}<5$~GeV (up to 1.06 for the Pb--Pb reference), with tilts between $+0.036$ and $+0.046$ for the nominal settings and between $-0.03$ and $+0.05$ over all variations. The slight hardening reflects the higher mean multiplicity of Ne--Ne collisions; multiplicity and density matching differ by less than 0.01 in the tilt, because the two systems have the same density at equal multiplicity. The minimum-bias Ne--Ne $\lam/\kzs$ is therefore predicted to agree with O--O within a few percent [Fig.~\ref{fig:nene}(b)]. This is a consistency test of the picture rather than a discriminating one: Ne--Ne should show the same tension with p--Pb at equal multiplicity and the same agreement at equal $\rho$ as O--O.

We generated $8.99\times10^5$ minimum-bias Ne--Ne events with EPOS4, using generator input for the 20+20 system provided by the EPOS authors (K.~Werner, private communication), and $5\times10^6$ events for each Angantyr variant (Sec.~S6). In EPOS4 the Ne--Ne/O--O ratio of $\lam/\kzs$ rises from 0.97 at 0.9~GeV to 1.03--1.07 above 2.5~GeV [Fig.~\ref{fig:nene}(c)], with a tilt of $+0.050\pm0.006$, consistent with the data-driven prediction ($+0.040\pm0.006$ without the hadronic cascade); in Angantyr the ratio is flat (tilts $+0.002$ to $+0.005$, $\pm0.003$). Within EPOS4, the p--Pb-based predictions of the Ne--Ne ratio repeat the O--O pattern: $s=+0.283\pm0.017$ matched in $\ntrk$, $+0.107\pm0.007$ in $\rho$ with nucleon sources, and $-0.040\pm0.008$ with constituents (O--O: $-0.020\pm0.009$), whereas Angantyr again shows no ordering (Table~S2).

\section{Discussion and conclusions}
\label{sec:discussion}

A possible reason why the transverse density orders the $\lam/\kzs$ shape better than multiplicity is the build-up of radial flow. In hydrodynamics, the radial boost is generated by pressure gradients, and the mean transverse momentum is, to good approximation, a function of the entropy density $s\propto(\mathrm{d}N/\mathrm{d}y)/\sperp$~\cite{Gardim:2019xjs}, so that the compact p--Pb system at high multiplicity expands more violently than the larger O--O or peripheral Pb--Pb system~\cite{Bozek:2013df,Bozek:2013ska}. Because the heavier $\lam$ is pushed further in $p_\mathrm{T}$ than the $\kzs$, the $\lam/\kzs$ ratio is a species-differential probe of this difference, complementary to the mean transverse momenta~\cite{ALICE:2013rdo,Pucillo:2026wdz}. A similar ordering is expected in core--corona approaches~\cite{Werner:2007bf,Becattini:2008ya,Aichelin:2008mi,Kanakubo:2019ogh,Ito:2026hec}. Our Glauber density is a proxy for both pictures, which the present data cannot distinguish; EPOS4, which combines both, reproduces the ordering.

The conclusions are subject to several limitations. (i)~The transverse density is computed with a Glauber model. With nucleons as sources, a tilt of $+0.14\pm0.04$ remains, and only with sub-nucleonic structure does it vanish; the full agreement is therefore conditional on a compact p--Pb geometry, which is favored by other studies~\cite{Welsh:2016siu,Mantysaari:2016ykx,Moreland:2018gsh} but not determined by the present analysis. (ii)~The $p_\mathrm{T}$ correlations of the systematic uncertainties had to be modeled; the multiplicity-matched p--Pb prediction is disfavored for all correlation models, but the significance of the residual deviations of the density-matched predictions depends on the model. (iii)~A residual energy dependence at high multiplicity cannot be excluded, but even one as large as the full measured 2.76--5.02~TeV change in Pb--Pb would shift the tilts by about 0.03, below their uncertainties of $\pm0.04$ and an order of magnitude below the multiplicity-matched p--Pb tilt; ALICE finds the multiplicity evolution of the $p_\mathrm{T}$-differential ratios in pp collisions to be similar at 7 and 13~TeV~\cite{ALICE:2019avo,ALICE:2020nkc}. (iv)~Only the minimum-bias O--O ratio has been measured, which averages over a wide range of multiplicities and relies on the interpolation and, in part, extrapolation of the reference classes. (v)~Only the $p_\mathrm{T}$ shape is tested.

In summary, using only published data, we find that the $p_\mathrm{T}$ shape of the $\lam/\kzs$ ratio in minimum-bias O--O collisions at $\sqrt{s_\mathrm{NN}}=5.36$~TeV is not reproduced by p--Pb collisions of the same multiplicity (tilt $+0.35\pm0.04$), but is consistent with that of peripheral Pb--Pb collisions. In a Glauber model calibrated on O--O data, O--O and peripheral Pb--Pb collisions of equal multiplicity have similar transverse densities, about half of that in p--Pb. Matching the p--Pb reference in density reduces the discrepancy by about 60\% with nucleons as sources; it is fully removed only when sub-nucleonic structure is included, so that this result is conditional on a compact geometry of high-multiplicity p--Pb collisions. The published radial-flow velocities of p--Pb and Pb--Pb collisions are close to each other at equal density. Within the uncertainties of the data and of the Glauber modeling, the transverse density therefore orders the $p_\mathrm{T}$ shape of $\lam/\kzs$ in these systems better than multiplicity does. EPOS4 shows the same pattern, already before the hadronic cascade, whereas PYTHIA Angantyr shows no dependence on the transverse geometry. For Ne--Ne collisions, the Glauber density at equal multiplicity is that of O--O, and the minimum-bias $\lam/\kzs$ ratio is predicted to agree with the O--O ratio within a few percent, slightly harder at high $p_\mathrm{T}$, as EPOS4 also finds. Centrality-differential measurements in O--O and Ne--Ne collisions will test the density ordering class by class and constrain the initial-state geometry of high-multiplicity p--Pb collisions. More generally, comparisons of small and large systems ``when compared at the same multiplicity'' of $p_\mathrm{T}$-differential observables that are sensitive to radial flow may need to account for their different transverse sizes; multiplicity remains a good ordering variable for $p_\mathrm{T}$-integrated strangeness enhancement~\cite{ALICE:2016fzo}.

\section*{Data availability}
All experimental inputs are public HEPData records, except the pp class fractions taken from Table~1 of Ref.~\cite{CMS:2016zzh}. The analysis code is available from the corresponding author upon reasonable request. Supplementary material (Secs.~S1--S6, Figs.~S1--S2, Tables~S1--S2) accompanies this article.

\section*{Funding}
This research did not receive any specific grant from funding agencies in the public, commercial, or not-for-profit sectors.

\section*{Declaration of competing interest}
The authors declare that they have no known competing financial interests or personal relationships that could have appeared to influence the work reported in this paper.

\section*{Acknowledgments}
We thank the CMS and ALICE Collaborations for making their results available on HEPData, and the EPOS authors, in particular K.~Werner, for providing the EPOS4 input for the Ne--Ne system and for advice.

\bibliographystyle{elsarticle-num}
\bibliography{references}

\end{document}


\maketitle

Section, equation, figure and table numbers without the prefix ``S'' refer to the Letter.

\section{Inputs, conventions, calibration and energy dependence}
\label{sec:s1}
All inputs are taken from HEPData:
\begin{itemize}
\item O--O at $\sqrt{s_\mathrm{NN}}=5.36$~TeV, minimum bias: $\kzs$ and $\lam$ spectra and their ratio at $|y|<1$, and the pp reference at the same energy~\cite{CMS:2026aie}; $\dndeta$ at $|\eta|<0.5$ in 5\% centrality classes from 0 to 70\%~\cite{CMS:2026sai}.
\item p--Pb at 5.02~TeV, pp at 7~TeV and Pb--Pb at 2.76~TeV: $\kzs$ and $\lam+\bar\lam$ spectra at $|y_\mathrm{cm}|<1$ in classes of the number of reconstructed tracks $\ntrk$ ($|\eta|<2.4$, $p_\mathrm{T}>0.4$~GeV), and the blast-wave parameters obtained by CMS~\cite{CMS:2016zzh}.
\item pp at 5.02~TeV~\cite{CMS:2019isl}; ALICE Pb--Pb at 2.76 and 5.02~TeV~\cite{ALICE:2013cdo,ALICE:2025cqy}, used for the collision-energy correction; ALICE pp $\dndeta$ at 5.02 and 13~TeV~\cite{ALICE:2022kol}, used for the energy dependence of the pp multiplicity in the calibration.
\end{itemize}
The only inputs not taken from HEPData are the fractions of the minimum-bias sample in the CMS pp multiplicity classes (0.93, 0.06, $6\times10^{-3}$, \dots), given in Table~1 of Ref.~\cite{CMS:2016zzh}. The CMS O--O measurement reconstructs $\lam$ only (no $\bar\lam$); the $\lam+\bar\lam$ spectra of the other analyses are divided by two. All $\lam$ yields are corrected for feed-down from $\Xi$ and $\Omega$ decays. Comparisons across experiments are sensitive to conventions (rapidity versus pseudorapidity normalization changes $\lam/\kzs$ by up to 60\% at 0.7~GeV), so the central analysis uses only CMS data.

\paragraph{Calibration.} The reference spectra come from a different CMS analysis (2016) than the O--O data (2026), and the pp reference classes are at 7~TeV. Both analyses measured minimum-bias pp collisions, which is the basis of the calibration of Eq.~(2). Because a minimum-bias ratio is a yield-weighted average, it is not equivalent to the reference taken at the mean multiplicity; the denominator of Eq.~(2) is therefore built from the six pp classes with their event fractions. The class multiplicities are scaled by $r=N(5.36~\mathrm{TeV})/N(7~\mathrm{TeV})=0.93$, so that the calibration does not absorb the genuine multiplicity difference between the two energies. Omitting this scaling ($r=1$) changes the multiplicity-matched p--Pb tilt from $+0.352$ to $+0.334$; without any calibration it is $+0.405$.

\paragraph{Pb--Pb energy correction.} The change of the $\lam/\kzs$ shape between 2.76 and 5.02~TeV at matched $\dndeta$ is taken from ALICE~\cite{ALICE:2013cdo,ALICE:2025cqy} and averaged over $\dndeta\approx55$ and 205. Normalized to unit mean, this factor deviates from unity by at most 6.5\% (rms 3.4\% over the $p_\mathrm{T}$ range); for $\dndeta\approx205$ alone it reaches 9\% (rms 3.9\%), for $\dndeta\approx55$ 6.1\% (rms 3.2\%). Applied to the multiplicity-matched Pb--Pb prediction it changes the tilt from $-0.014$ to $+0.014$.

\paragraph{5.02 to 5.36~TeV.} The remaining step from 5.02 to 5.36~TeV, which also separates the p--Pb reference from the O--O data, is 0.11 of the 2.76--5.02~TeV step in $\ln\sqrt{s_\mathrm{NN}}$. Scaling the measured factor accordingly changes the tilts by less than 0.005: p--Pb matched in $\ntrk$ $+0.352\to+0.357$, in $\rho$ with nucleons $+0.138\to+0.142$ and with constituents $-0.021\to-0.016$; Pb--Pb $+0.014\to+0.017$. At matched multiplicity, the multiplicity evolution of the $p_\mathrm{T}$-differential ratios in pp collisions at 13~TeV is similar to that at 7~TeV, and the $p_\mathrm{T}$-integrated ratios agree at similar multiplicities~\cite{ALICE:2019avo,ALICE:2020nkc}, which suggests that the energy dependence at fixed multiplicity is small.

\paragraph{Extrapolated classes.} With the saturated shape beyond the highest p--Pb class the full multiplicity-matched tilt is $+0.33$; for $\kappa=2.40$ (3.10), with two (four) extrapolated classes carrying 30\% (52\%) of the charged-particle yield, it is $+0.32$ ($+0.38$).

\section{The pp-based prediction}
\label{sec:s2}
\begin{figure}[h]
\centering
\includegraphics[width=0.55\textwidth]{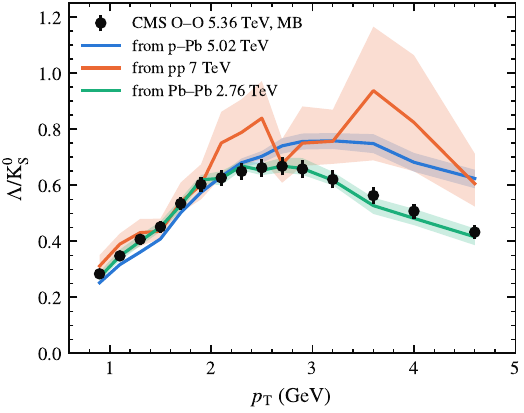}
\caption{As Fig.~1, with the prediction from the CMS 7~TeV pp classes added (best-fit scale 1.54).}
\label{fig:pp}
\end{figure}
Figure~\ref{fig:pp} adds the prediction based on the 7~TeV pp classes, which is shown for completeness but not used in the tests. The CMS pp classes extend only to $\ntrk\approx137$, whereas the most central O--O classes correspond to $\ntrk\approx370$: six of the twenty O--O classes, which carry two thirds of the charged-particle yield, lie beyond the measured pp range. There the prediction is an extrapolation in multiplicity by up to a factor of 2.7, which amplifies the 2--4\% statistical uncertainties of the highest pp classes to 6--17\% on the minimum-bias prediction (for p--Pb and Pb--Pb, whose classes reach $\ntrk\approx230$, it is 0.6--2\% and 1--5\%). In addition, the calibration to the 5.36~TeV pp measurement corrects the energy difference at minimum-bias pp multiplicity, but not a possible energy dependence at high multiplicity.

\section{Covariance models and systematic variations}
\label{sec:s3}
HEPData provides only total systematic uncertainties. Each relative systematic uncertainty is therefore decomposed into a normalization-like component (fully correlated in $p_\mathrm{T}$), a component linear in $\log p_\mathrm{T}$ (tilt), a component restricted to low $p_\mathrm{T}$, and an uncorrelated remainder. Results are given for several decompositions, from fully uncorrelated to almost fully correlated; each component is fully correlated in $p_\mathrm{T}$ except the last, and statistical uncertainties are added on the diagonal. The nominal ``balanced'' decomposition assigns variance fractions of 0.3, 0.3, 0.2 and 0.2 to the normalization, tilt, low-$p_\mathrm{T}$ and uncorrelated components.

Table~\ref{tab:syst} lists the variations considered for the two key tilts. The experimental uncertainties of the O--O and reference ratios enter through the covariance model; the remaining entries are variations of the procedure. The multiplicity-matched tilt stays above $+0.29$ in every variation, and the density-matched tilt with constituent sources stays within $\pm0.04$ of zero. The density-matched tilt depends mainly on how the transverse size of high-multiplicity p--Pb collisions is modeled, which is the dominant uncertainty of the density picture.

The covariance model uses the uncertainties of the reference ratio in one class. To include the statistical fluctuations of all reference classes as they propagate through the interpolation, and beyond the highest class through the extrapolation, we refit the tilts on 400 bootstrap replicas in which every reference spectrum is fluctuated within its statistical uncertainty. The spread of the refitted tilts is 0.007--0.009 for the p--Pb-based predictions and 0.008--0.015 for the Pb--Pb-based one, with no shift of the central values; added in quadrature, it changes the uncertainties of Table~1 by at most 0.003.

\begin{table}[h]
\caption{Tilt $s$ of the p--Pb-based predictions under variations of the analysis. ``Constituents'' denotes density matching with three constituents per nucleon ($\sigma_c=0.4$~fm) unless stated otherwise.}
\label{tab:syst}
\centering
\begin{tabular}{lcc}
\toprule
Variation & $\ntrk$-matched & $\rho$-matched \\
\midrule
Nominal (balanced covariance) & $+0.352\pm0.040$ & $-0.021\pm0.040$ (constituents)\\
Uncorrelated uncertainties & $+0.329\pm0.052$ & $-0.035\pm0.052$ (constituents)\\
$\kappa$, extrapolation, calibration & $+0.29$ \dots\ $+0.43$ & $-0.04$ \dots\ $+0.03$ (constituents)\\
Constituent spread $\sigma_c=0.55$~fm & --- & $-0.036\pm0.040$\\
Nucleon sources, $w=0.5$~fm, $f=0.5$ & --- & $+0.138\pm0.040$\\
Nucleon sources, $w=0.4$~fm & --- & $+0.097\pm0.040$\\
Nucleon sources, $f=0.4$ & --- & $+0.132\pm0.040$\\
\bottomrule
\end{tabular}
\end{table}

\section{Glauber model}
\label{sec:s4}
The Monte Carlo Glauber model~\cite{Miller:2007ri,Loizides:2017ack} uses Woods--Saxon $^{208}$Pb, harmonic-oscillator $^{16}$O and $\sigma_\mathrm{NN}=61.8$, 67.6 and 68.2~mb at 2.76, 5.02 and 5.36~TeV. Event classes are defined with estimators that mimic the experimental ones; the model reproduces the mean $\ntrk$ of the CMS Pb--Pb classes and those of p--Pb to within 8\%. In the constituent version~\cite{Loizides:2016djv} each nucleon is replaced by three constituents with Gaussian spread $\sigma_c=0.4$ or 0.55~fm, the constituent--constituent cross section (23.4 or 15.6~mb at 5.02~TeV) is tuned to reproduce $\sigma_\mathrm{NN}$, and the O--O centrality dependence of $\dndeta$ is then best described with $f=0.8$ (1.5--2.2\% rms). It reproduces $\sigma_\mathrm{inel}=2.09$~b for p--Pb and 1.35~b for O--O. Our own blast-wave implementation~\cite{Schnedermann:1993ws} reproduces the CMS parameters of Fig.~2(a) when fitted with statistical uncertainties only, as in Ref.~\cite{CMS:2016zzh}.

\paragraph{Ne--Ne.} $^{20}$Ne is implemented as a deformed Woods--Saxon distribution with random orientation, with the parameters of TGlauberMC~v3.3~\cite{Loizides:2025ule} ($R=2.724$~fm, $a=0.498$~fm, $\beta_2=0.490$, $\beta_3=0.216$, $\beta_4=0.306$); a more strongly deformed nucleus ($\beta_2=0.721$) and a spherical two-parameter Fermi distribution are used as variations. Since Ne--Ne and O--O are measured at the same energy, the per-source yield is taken from the O--O calibration, so that the Ne--Ne centrality dependence of $\dndeta$ is a prediction. We test it against the Ne--Ne/O--O ratio of $\dndeta$ measured by ATLAS in 5\% centrality classes from 0 to 80\%~\cite{ATLAS:2026zgq}, in which most of the centrality-determination and normalization uncertainties cancel. For the nominal nucleus the model exceeds the measured ratio by 3\% on average (rms deviation 3.4\%, 5.7\% in the 0--5\% class) with nucleon sources and by 4.5\% with three constituents; the more deformed nucleus lies 3--4\% below the data, and the spherical nucleus agrees best (rms 2.5\% and 1.7\%). The two deformed parameterizations thus bracket the data. The ATLAS and CMS O--O measurements themselves differ by $-5\%$ (0--5\%) to $+27\%$ (65--70\%), so absolute values from the two experiments are not combined: the Ne--Ne class multiplicities are the CMS O--O values multiplied by the ATLAS Ne--Ne/O--O ratio. The mean minimum-bias $\dndeta$ is then 47.0 for Ne--Ne and 39.8 for O--O with the same construction.

\section{Generators}
\label{sec:s5}
EPOS~4.0.3~\cite{Werner:2023zvo} is run in its full configuration (core--corona separation, 3+1D viscous hydrodynamics for the core, UrQMD~\cite{Bass:1998ca,Bleicher:1999xi}), with ten particlization samplings per hydrodynamic evolution. PYTHIA~8.317 Angantyr~\cite{Bierlich:2018xfw,Bierlich:2022pfr} is run with the default Monash settings, with rope hadronization and string shoving~\cite{Bierlich:2014xba,Bierlich:2017vhg} (untuned, as stated in the PYTHIA manual), and with QCD-based color reconnection. We generated $9\times10^5$ minimum-bias O--O events at 5.36~TeV and $9\times10^5$ p--Pb events at 5.02~TeV (half of them restricted to $b<4$~fm to populate high multiplicities) with EPOS, and $5\times10^6$ events per system and variant with Angantyr. Acceptance ($|y_\mathrm{cm}|<1$, $\lam$ only, $\Xi$ and $\Omega$ stable so that no weak feed-down enters) and multiplicity estimators (forward transverse energy for O--O, the laboratory-frame $\ntrk$ for p--Pb) follow the data. Tilt uncertainties are obtained by bootstrap over independent hydrodynamic evolutions or event blocks. Reference classes are interpolated by a weighted quadratic fit in $\log\ntrk$ or $\log\rho$; linear interpolation changes the tilts by less than 0.02. With the larger constituent spread the EPOS4 density-matched tilt is $-0.040\pm0.009$.

Three of the twenty O--O classes in EPOS (two in Angantyr) lie beyond the most central p--Pb class with sufficient statistics, so the multiplicity-matched prediction is extrapolated there and fluctuates above $p_\mathrm{T}\approx2.8$~GeV. Restricting the tilt fit to $p_\mathrm{T}<2.8$~GeV changes the EPOS values to $+0.20$ (multiplicity) and $-0.04$ (density, constituent sources); the conclusions are unchanged. The Angantyr variant with QCD-based color reconnection gives tilts consistent with zero for all matchings ($+0.012\pm0.009$ for multiplicity, $-0.012\pm0.004$ for density matching).

\begin{figure}[h]
\centering
\includegraphics[width=0.55\textwidth]{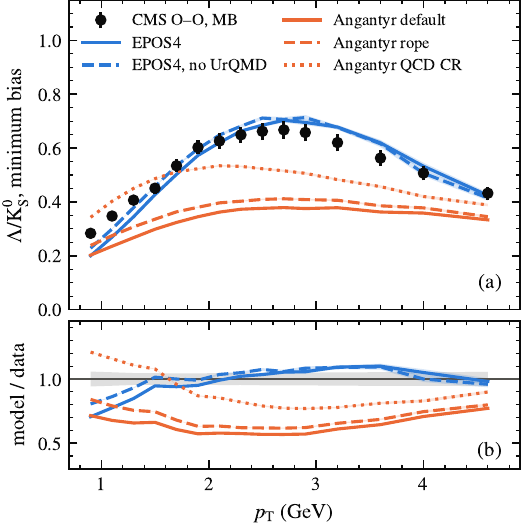}
\caption{(a) Minimum-bias O--O $\lam/\kzs$ measured by CMS~\cite{CMS:2026aie} compared with EPOS4 (blue; solid: full model, dashed: without the hadronic cascade) and PYTHIA Angantyr (orange; solid: default, dashed: rope hadronization and string shoving, dotted: QCD-based color reconnection). Bands: statistical uncertainties of the models. (b) Ratio to the data; gray band: data uncertainty. No scale factor is applied, since the generators predict the absolute ratio. Shape test with uncorrelated data uncertainties (best-fit scale $a$; $\chi^2/\mathrm{ndf}$): EPOS4 (1.02; 67/14), EPOS4 without UrQMD (0.99; 34/14), Angantyr default (1.59; 54/14), Angantyr rope (1.44; 65/14), Angantyr QCD CR (1.07; 143/14).}
\label{fig:models}
\end{figure}

\paragraph{Absolute comparison.} In absolute terms neither model describes the measured O--O ratio in detail (Fig.~\ref{fig:models}). EPOS4 reproduces its normalization (best-fit scale 1.02) and the position of the maximum, but underestimates the ratio below 1.5~GeV by up to 30\% and overestimates it by about 10\% at 3--3.6~GeV (shape $\chi^2/\mathrm{ndf}=67/14$ with uncorrelated data uncertainties). Without the hadronic cascade the low-$p_\mathrm{T}$ part is closer to the data (34/14). Default Angantyr underestimates the ratio by about 40\% at the maximum and has almost no intermediate-$p_\mathrm{T}$ enhancement (54/14); rope hadronization raises it by only about 10\% (65/14). The QCD color-reconnection variant gives the poorest description of the measured shape ($\chi^2/\mathrm{ndf}=143/14$): it overestimates the ratio below 1.5~GeV and underestimates it by about 20\% above 2.5~GeV.

\section{Ne--Ne generator results}
\label{sec:s6}
The EPOS4 Ne--Ne sample ($8.99\times10^5$ minimum-bias events at 5.36~TeV, configuration of Sec.~\ref{sec:s5}) uses generator input for the 20+20 system provided by the EPOS authors (K.~Werner, private communication); two of 1800 jobs, which stopped inside the generator, were discarded. Angantyr: $5\times10^6$ events for each of the three variants. Both generators describe $^{20}$Ne as a spherical Woods--Saxon nucleus. Both overestimate the Ne--Ne/O--O ratio of $\dndeta$ in the 0--5\% class (1.27 for EPOS4 and all Angantyr variants, ATLAS 1.21), similar to our Glauber model with the deformed nucleus. In Angantyr the Ne--Ne/O--O ratio of $\lam/\kzs$ is flat, consistent with the weak multiplicity dependence of its $\lam/\kzs$ shape; the slight hardening in EPOS4 is consistent with the higher Ne--Ne multiplicity combined with a multiplicity-dependent radial boost, which also underlies the data-driven prediction. Table~\ref{tab:nene} summarizes the results.

\begin{table}[h]
\caption{Ne--Ne results in EPOS4 ($8.99\times10^5$ Ne--Ne and $9\times10^5$ O--O events) and PYTHIA Angantyr ($5\times10^6$ events per system and variant); statistical uncertainties. Second column: Ne--Ne/O--O ratio of $\dndeta$ in the 0--5\% class (ATLAS: 1.213~\cite{ATLAS:2026zgq}). Third column: tilt of the Ne--Ne/O--O ratio of minimum-bias $\lam/\kzs$ (data-driven prediction: $+0.036$ to $+0.046$ for the nominal settings). Last two columns: tilt of the model's p--Pb-based prediction of its own Ne--Ne ratio, matched in $\ntrk$ or in $\rho$ (three constituents); O--O values in parentheses.}
\label{tab:nene}
\centering
\begin{tabular}{lcccc}
\toprule
Model & Ne/O $\dndeta$ & Ne/O tilt & $s$, $\ntrk$ & $s$, $\rho$ \\
\midrule
EPOS4   & 1.268 & $+0.050(6)$ & $+0.283(17)$ & $-0.040(8)$ \\
        &       &             & $(+0.274)$   & $(-0.020)$ \\
EPOS4, no UrQMD & 1.268 & $+0.040(6)$ & $+0.326(18)$ & $+0.034(8)$ \\
        &       &             & $(+0.312)$   & $(+0.049)$ \\
Angantyr default & 1.272 & $+0.005(3)$ & $-0.087(11)$ & $-0.024(4)$ \\
        &       &             & $(-0.064)$   & $(-0.022)$ \\
Angantyr rope & 1.269 & $+0.003(3)$ & $-0.086(8)$  & $-0.032(4)$ \\
        &       &             & $(-0.070)$   & $(-0.032)$ \\
Angantyr QCD CR & 1.272 & $+0.002(3)$ & $+0.013(10)$ & $-0.007(4)$ \\
        &       &             & $(+0.012)$   & $(-0.012)$ \\
\bottomrule
\end{tabular}
\end{table}

\bibliographystyle{elsarticle-num}
\bibliography{references}